\documentclass[aps, floatfix, prd, twocolumn, 10pt, nofootinbib]{revtex4-1}
\usepackage{amsmath, amsthm, amssymb}
\usepackage{epsfig}
\usepackage{slashed}

\graphicspath{{graphics/}}

\newcommand{\eq}[1]{Eq.~(\ref{#1})}
\newcommand{\eqs}[1]{Eqs.~(\ref{#1})}
\newcommand{\fig}[1]{Fig.~\ref{#1}}

\begin{document}

 \title{The Quark Propagator in the NJL Model
in a self-consistent $1/N_c$ Expansion}
 \date{\today}
 \author{D. M\"uller} \author{M. Buballa} \author{J. Wambach}
 \affiliation{Institut f\"ur Kernphysik, Technische Universit\"at Darmstadt, Germany}

\begin{abstract}
The quark propagator is calculated in the Nambu-Jona-Lasinio (NJL) model in a 
self-consistent $1/N_c$-expansion at next-to-leading order. The calculations are 
carried out iteratively in Euclidean space. The chiral quark condensate and its 
dependence on temperature and chemical potential is calculated directly and compared 
with the mean-field results. 
In the chiral limit, we find a second-order phase transition at finite temperature 
and zero chemical potential, in agreement with universality arguments. 
At zero temperature and finite chemical potential, the phase transition is first 
order. 
In comparison with the mean-field results, the critical temperature and chemical 
potential are slightly reduced.
We determine spectral functions from the Euclidean propagators
by employing the Maximum-Entropy-Method (MEM). 
Thereby quark and meson masses are estimated and decay channels identified. 
For testing this method, we also apply it to evaluate perturbative spectral 
functions, which can be calculated directly in Minkowski space. 
In most cases we find that MEM is able to reproduce the rough features of the 
spectral functions, but not the details.

\end{abstract}

\maketitle

\section{Introduction}
Describing the low-energy sector of QCD is a complicated task due to its 
strong coupling. Especially investigating the region of the chiral and 
deconfinement phase transition requires non-perturbative methods. 
To describe these phenomena, one therefore often employs effective 
models, which give a simplified description of the theory and are valid in a 
limited energy range. In this article, we focus on the Nambu-Jona-Lasinio 
(NJL) model \cite{NJL61}, where the quark-gluon interactions of QCD 
are substituted by effective 4-quark interactions. This model does not exhibit 
confinement, which is its major lack, but it incorporates chiral symmetry. 
Spontaneous and explicit breaking of this symmetry as well as its restoration 
at high temperatures or densities can be realized.

Despite the simplifications in the interaction, the NJL model cannot be solved 
exactly, but further approximations are necessary.\footnote{Strictly speaking,
since the NJL model is non-renormalizable, a unique exact solution does not 
even exist, but the results depend on the regularization scheme. 
An interesting alternative to the continuum methods discussed in 
this article is to solve the NJL model on the lattice \cite{Hands:2002mr}.
In this case the non-renormalizability has the consequence that
there is no continuum limit, so that the results depend on the choice of the 
lattice.} 
In most publications, the model has been treated in mean-field
(Hartree) and random-phase approximation to describe the chiral dynamics 
of quarks and the meson spectrum, both, in vacuum and in hot and dense matter 
\cite{VW91, Klev92, HK94}. 
A shortcoming of these approximations is that the effects of mesons on the 
quark propagator are not included. In the quark sector this leads to the 
wrong universal behavior at the phase transition and produces delta peaked 
quark spectral functions.
As a consequence, for instance, they are not suitable to be used in the
Kubo formula for the calculation of shear viscosities~\cite{Iwasaki:2006dr}.
In the mesonic sector, hadronic decay channels, like 
$\rho \rightarrow \pi\pi$, are not included, which are the physical
decay channels in the confined phase. 
 
These processes can be included systematically within a $1/N_c$-expansion, 
i.e., within an expansion in the inverse number of colors, beyond the
leading order, see, e.g., 
Refs.~\cite{QK94,ENV94,DSTL95,Zhu95,BKRSV95,NBC+96,FB96,OBW00,Oer00,BBRV08}. 
Here one can basically follow two different approaches: 
In the ``strict $1/N_c$ expansion scheme'' one first solves the gap equation 
in leading order, i.e., Hartree approximation, and then adds the $1/N_c$
corrections to the quantities of interest perturbatively, without 
modifying and solving the gap equation again. This method yields
good results, e.g., for the rho-meson dominated electromagnetic pion form 
factor in the time-like region~\cite{OBW00}  
as well as for the low-temperature behavior of the 
quark condensate~\cite{BKRSV95,Oer00} and the pressure~\cite{BBRV08}. Above 
the critical temperature the model has been studied in \cite{Kitazawa:2005mp}
where the influence of soft modes on the quark spectral function was investigated.
On the other hand, the perturbative approach breaks down in the vicinity
of the phase transition, where a method is needed  
which incorporates the  $1/N_c$ corrections self-consistently in the
gap equation.
First attempts in this direction have been performed in 
Refs.~\cite{DSTL95,FB96,Oer00},
but in a simplified approach, where nonlocal contributions to the quark 
self-energy have been neglected. This scheme is thermodynamically 
inconsistent and it was found that the chiral phase transition at finite 
temperature is first order for two quark 
flavors~\cite{FB96,Oer00}, in contradiction to universality 
arguments~\cite{PW84}.

In the present paper, we derive a self-consistent solution of the gap 
equation using the so-called $\Phi$-derivable theory 
in next-to-leading order without further approximations.
This approach is thermodynamically consistent and allows a
meaningful investigation of the phase transition. 
Due to the nonlocal self-energy contributions,
the structure of the equations is rather involved and we use 
the imaginary time (Matsubara) formalism to keep them on a tractable level.
This allows us to study static properties, like the quark condensate
at zero and finite temperature. 
On the other hand, the analytic continuation of dynamical quantities
to real times is problematic. In order to gain information about quark 
and meson spectral functions in Minkowski space, 
we therefore employ the Maximum-Entropy Method (MEM).
 
The remainder of this article is organized as follows.
In Sec.~\ref{model} we summarize the basics of the model and the 
leading-order formalism. In Sec.~\ref{Phi} we introduce the
$\Phi$-derivable theory and formally derive the gap equation for the 
quark propagator in next-to-leading order (NLO) in $1/N_c$. 
The numerical results for the self-consistent solutions in the Matsubara
formalism are shown in Sec.~\ref{results}. There, we also discuss
the behavior of the quark condensate as a function of temperature
and chemical potential.
In Sec.~\ref{MEM} we review the basic ideas of MEM and apply them
to study quark and meson spectral functions.
We conclude with a summary in Sec.~\ref{summary}.

\section{Model and leading order formalism}\label{model}

We use a two-flavor NJL model with a scalar and pseudoscalar interaction, 
given by the Lagrangian
\begin{equation}
 \mathcal{L}
 \;=\;
 \bar{q}(i\slashed{\partial}-m_0)q
 +G[(\bar{q}q)^2+(\bar{q}i\gamma_5\vec{\tau}q)^2]
\end{equation}
with a dimensionful coupling constant $G$ and the Pauli matrices $\vec\tau$ 
in isospin space. $m_0$ is a small bare quark mass, which explicitly breaks 
chiral symmetry. For calculations in the chiral limit it is set to zero.
In this limit the Lagrangian is invariant under $SU(2)_L\times SU(2)_R$
transformations. For nonvanishing but small values of $m_0$, this is
still an approximate symmetry of the model. 

The $1/N_c$ counting scheme is introduced in the NJL model by assuming
that the quark fields $q$ have $N_c$ color degrees of freedom.
Consequently, a closed quark loop yields a factor $N_c$.
Furthermore, it is assumed that the coupling constant $G$ scales like 
$1/N_c$. In this article these rules are used to organize the diagrams 
in a systematic and symmetry conserving way. In all explicit
calculations, however, we take the physical number of colors, $N_c=3$.

\begin{figure}[t]
 \includegraphics[width=0.45\textwidth]{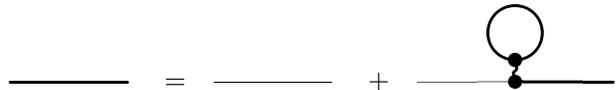}
\caption{Dyson-equation for the Hartree quark propagator (bold lines). 
Thin lines represent the bare propagator, the wavy line the bare interaction.}
\label{dyson_hartree}
\end{figure}

In any approximation the full quark propagator $S(k)$ is given by
\begin{equation}
	S^{-1}(k) \;=\; S_0^{-1}(k) - \Sigma(k)
\end{equation}
with the inverse bare propagator $S_0^{-1}(k) = \slashed k - m_0$ and a 
self-energy $\Sigma(k)$.
For large enough couplings, chiral symmetry is spontaneously broken and the
quarks acquire a dynamical mass much larger than the bare mass.
In leading order in $1/N_c$ this is described by the Hartree Dyson-equation, 
\fig{dyson_hartree}. 
In the self-energy insertion on the right-hand side (r.h.s.), 
the factor $N_c$ of the quark loop is compensated by a factor $1/N_c$ 
from the vertex.
Hence, if we assume that the bare propagator is of the order $N_c^0$, 
we find that the dressed propagator in Hartree approximation is strictly
of the  order $N_c^0$ as well.
Note that in this paper we draw the local four-point interaction as a wavy 
line, indicating the direction of the interaction. In the present example 
this prevents confusion of the Hartree term with the Fock term, which is
suppressed by one order of $1/N_c$.

In vacuum, the Hartree self-energy is given by
\begin{equation}
 \Sigma_H  
 \;=\; 
  2iG \int\frac{d^4k}{(2\pi)^4}\mbox{Tr}(S(k))\,.
\label{SigmaH}
\end{equation}
It is local and purely scalar and therefore corresponds to a constant 
shift in the quark mass,
\begin{equation}
 m_H \;=\;m_0 + \Sigma_H\,. 
\end{equation}
The dressed or ``constituent quark'' mass $m_H$ is the scalar part of the 
inverse Hartree propagator $S^{-1}(k)\;=\;\slashed{k}-m_H$.
In this way it enters the r.h.s. of \eq{SigmaH}, thereby 
giving rise to a self-consistency problem. 

To describe the system at nonvanishing temperature $T$ and chemical
potential $\mu$, we apply the Matsubara formalism.
The quark propagator is then defined at discrete imaginary energies
$i\omega_n + \mu$, with fermionic Matsubara frequencies
$\omega_n = (2n+1)\pi T$. 
Accordingly, the energy integration in fermionic loop integrals is 
replaced by a sum,
\begin{equation}
    i \int\frac{d^4k}{(2\pi)^4} f(k_0,\vec k)
    \;\rightarrow\;
    -T\sum_n\int\frac{d^3k}{(2\pi)^3} f(i\omega_n+\mu,\vec k)\,.
\label{replace}
\end{equation}
In the Hartree approximation, where the self-energy is just a
constant, the analytic continuation of the propagator to real energies 
is, of course, trivial. 
However, this will no longer be the case at NLO. 
\begin{figure}[b]
 \includegraphics[width=0.45\textwidth]{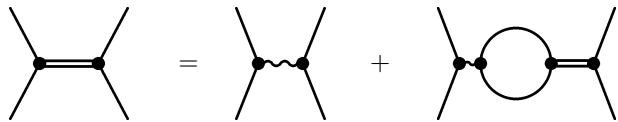}
\caption{Bethe-Salpeter-equation for quark-antiquark scattering. 
Double lines denote RPA meson propagators.}
\label{bse_rpa}
\end{figure}

Mesons are described by a Bethe-Salpeter-equation for the 
quark-antiquark $T$-matrix,
\begin{equation}
 i\hat T(q) \;=\; i\hat K + i\hat K(-i\hat\Pi(q))i\hat T(q)\,.
\label{BSE}
\end{equation}
The leading order corresponds to the random-phase approximation (RPA), 
depicted in \fig{bse_rpa}.
In this case
\begin{equation}
	i\hat K \;=\;2iG \sum_M (\Gamma_M\otimes\Gamma_M)
\end{equation}
is the bare scattering kernel and
\begin{alignat}{1}
J_{M}(q)
\;\equiv\;&\Gamma_M\hat\Pi(q)\Gamma_M 
\nonumber\\
\;=\;& i\int\frac{d^4k}{(2\pi)^4} \mbox{Tr}[\Gamma_M S(k+q) \Gamma_M S(k)]
\label{Jmes}
\end{alignat}
are the quark-antiquark polarization functions
in the scalar and pseudoscalar channels,
\begin{equation}
	\Gamma_s \;=\; 1\hspace{-0.13cm}1, ~~~ 
        \Gamma_{p,a} \;=\; i\gamma_5\tau_a\,.
\end{equation}
Here we have used that these channels do not mix. 
\eq{BSE} is then easily solved with the ansatz 
\begin{equation}
\hat T_M = -D_M(\Gamma_M\otimes\Gamma_M)\,, 
\end{equation}
which yields
\begin{equation}
 D_M(q)\;=\;\frac{-2G}{1-2GJ_{M}(q)}\,.
\end{equation}
Following Refs.~\cite{OBW00,Oer00}, we will call the functions
$D_M(q)$ ``meson propagators'' although they are not properly normalized. 
In particular, the meson masses are given by the pole positions of the
propagators,
\begin{equation}
\left.D^{-1}_M(q)\right\vert_{q^2=m_M^2} \;=\; 0\,.
\label{mM}
\end{equation}

When we expand the r.h.s. of \fig{bse_rpa} into a geometric series,
we see that the term with $n$ quark loops contains $n+1$ four-point vertices.
Thus, applying the $1/N_c$ counting rules, we find that the RPA meson 
propagators are strictly of the order $1/N_c$. 

In the Matsubara formalism, the meson propagators and polarization functions
are defined at discrete imaginary energies $i\omega_m$, 
with bosonic Matsubara frequencies $\omega_m = 2m\pi T$. 
The determination of meson masses according to \eq{mM} then requires
the analytic continuation of the propagator to real energies. 
Again, this is easily done in the Hartree $+$ RPA scheme, but will be 
non-trivial at NLO.

\section{$\Phi$-derivable theory}\label{Phi}

As motivated in the introduction we are aiming at a self-consistent 
extension of the approximation scheme beyond the leading order
in $1/N_c$.
To this end we apply the $1/N_c$-expansion on the level of the thermodynamic 
potential using the so-called $\Phi$-derivable theory \cite{LW60, BK61}. 
This scheme preserves all important symmetries.

The full thermodynamic potential is given by
\begin{equation}
\Omega[S] \;=\; i\,\mathbf{Tr}\ln S^{-1} + i\,\mathbf{Tr}(\Sigma S) + \Phi[S]\,,
\end{equation}
where $S$ and $\Sigma = S_0^{-1} - S^{-1}$ are the full quark propagator 
and the full self-energy, respectively, and $\mathbf{Tr}$ denotes a 
functional trace over all space-time and internal degrees of freedom. 
The functional $\Phi[S]$ summarizes all closed two-particle irreducible (2PI) 
diagrams~\cite{CJT74}.

The stationarity condition $\frac{\delta \Omega}{\delta (iS)} = 0$
implies that 
\begin{equation}\label{phis}
\Sigma(x) \;=\; -\frac{\delta \Phi}{\delta (iS(x))}\,,
\end{equation}
i.e., the self-energy can be obtained as a functional derivative of $\Phi$.
Diagrammatically, this corresponds to cutting a single quark line of
$\Phi$ at all possible places. 
In turn, $\Phi$ depends on $\Sigma$ via the full quark propagator.
\eq{phis} therefore constitutes a self-consistency problem.

Similarly, the symmetry conserving scattering kernel for the
mesonic BSE can be obtained as
\begin{equation}\label{phik}
	\hat K(x,y) \;=\; -\frac{\delta^2 \Phi}{\delta (iS(x))\delta (iS(y))}\,, 
\end{equation}
corresponding to cutting the $\Phi$-functional twice. 
However, unlike for the self-energy, this is not a self-consistency problem
because $\Phi$ does not depend on $\hat K$.
\begin{figure}[t]
 \includegraphics[width=0.15\textwidth]{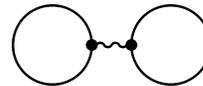}
\caption{Leading order contribution to $\Phi$.}
\label{glasses}
\end{figure}

Self-consistent approximation schemes can now be introduced by performing 
truncations of the functional $\Phi$.
In the present article, we expand $\Phi$ in powers of $1/N_c$ to
next-to-leading order. The leading-order contribution corresponds to the 
"glasses" diagram, shown in \fig{glasses}. 
In vacuum it is given by
\begin{equation}
\Phi^{(0)}[S] 
= 
-G \sum_M \left(-i\int\frac{d^4 k}{(2\pi)^4}
 \mbox{Tr}(\Gamma_M S(k))\right)^2\,,
\label{Phi0}
\end{equation}
which can be generalized in the Matsubara formalism by the replacement
(\ref{replace}).
As it contains two quark loops and one four-point vertex, it
is of the order $\mathcal{O}(N_c)$.
Cutting one or two quark lines, we reproduce our earlier result that
the Hartree self-energy and the RPA scattering kernel, respectively, 
are the corresponding leading-order expressions in this expansion.
Formally, this can also be obtained from \eqs{phis} and (\ref{phik}),
where one has to take into account that the transformation to
momentum space brings in extra factors of $(2\pi)^4$.
For instance, for the self-energy one gets
\begin{equation}
\Sigma(k) 
\;=\; 
i(2\pi)^4\frac{\delta \Phi}{\delta S(k)}\,,
\label{phisp}
\end{equation}
which, when applied to \eq{Phi0}, indeed yields \eq{SigmaH}.

The fact that the Hartree self-energy and the RPA scattering kernel
can consistently be derived from the same $\Phi$-functional guarantees 
that this approximation scheme is symmetry conserving. In particular
chiral Ward identities and low-energy theorems are fulfilled and the 
RPA pion is massless in the chiral limit, as required by Goldstone's theorem. 

\begin{figure}[t]
 \includegraphics[width=0.45\textwidth]{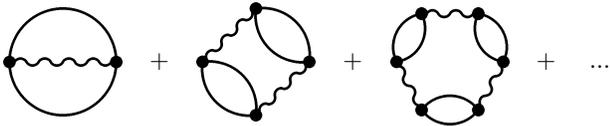}
\caption{NLO contribution to $\Phi$.}
\label{ringsum}
\end{figure}

\begin{figure}[b]
 \includegraphics[width=0.45\textwidth]{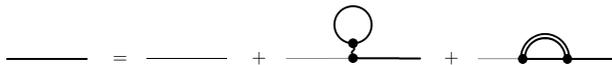}
\caption{NLO gap equation. The thin and bold lines represent bare
and dressed quark propagators, respectively.
The double line symbolizes RPA-like mesons as defined in \fig{bse_rpa}
but involving the self-consistent solution for the quark propagator. 
}\label{gap_nlo}
\end{figure}

The NLO correction to $\Phi$ is given by the ring sum,
depicted in \fig{ringsum}.
Taking into account the appropriate symmetry factors,
these diagrams can be combined to a logarithm. One obtains
\begin{equation}\label{phi_nlo}
\Phi^{(1)}[S] 
\;=\; 
-\frac{i}{2}\sum_M \int\frac{d^4 q}{(2\pi)^4}\mbox{ln}(1-2G J_M(q))\,,
\end{equation}
which depends on $S$ through the polarization function $J_M$, \eq{Jmes}.
Applying again \eq{phisp}, we find the following NLO correction to the 
self-energy:
\begin{equation}\label{self_nlo}
\Sigma^{(1)}(k) 
\;=\; 
i\sum_M \int\frac{d^4 q}{(2\pi)^4} D_M(q)\Gamma_M S(k-q)\Gamma_M\,.
\end{equation}
It describes the dressing of the quark propagator by an RPA meson
and corresponds to the insertion in the last diagram in \fig{gap_nlo}. 
Recalling that the RPA meson propagators are of the order $1/N_c$,
this self-energy term yields a correction of the order $1/N_c$ to the
quark propagator.
However, when the diagrams are iterated in the gap equation, 
as shown in the figure, higher orders are generated.
Therefore both, the self-consistent quark propagator and the individual 
self-energy contributions, are no longer of strict orders in $1/N_c$.
Obviously, the same is true for $\Phi^{(0)}$ and $\Phi^{(1)}$
which contain higher orders in $1/N_c$ as well, when the self-consistent 
quark propagator is used. In the present scheme,
the $1/N_c$ counting is thus introduced on the level of skeleton diagrams
for the $\Phi$-functional, i.e., before dressing the propagators. 

As we have discussed, at leading order RPA mesons together with the
Hartree gap equation are consistent with chiral Ward identities so that,
in the chiral limit, pions emerge as massless Goldstone bosons in that 
scheme.
In the same way one can construct mesons which are consistent with
the NLO gap equation. To that end one has to calculate the NLO corrections
to the scattering kernel by applying \eq{phik} to $\Phi^{(1)}$,
corresponding to  cutting the diagrams in \fig{ringsum} twice,
and iterate them together with the leading-order kernel in the BSE.
Again, this scheme preserves chiral symmetry and the resulting pions
are massless in the chiral limit.

In this context it should be noted that the mesons which enter the 
NLO self-energy diagram are {\it not} the NLO-corrected mesons,
but RPA mesons. To be precise, they are obtained from the BSE with
the leading-order scattering kernel, \fig{bse_rpa}, but the
polarization functions $J_M$, \eq{Jmes}, involve the self-consistent 
NLO quark propagators. As a consequence, these ``intermediate mesons'' 
are not restricted by chiral Ward identities and the pions are not
necessarily massless in the chiral limit. 
Formally, this problem is a higher-order effect in $1/N_c$ but, as we 
will see below, it is quite severe.
It can be avoided by performing a ``strict $1/N_c$-expansion'',
where only the Hartree gap equation is solved
self-consistently and the NLO corrections are added perturbatively,
discarding all higher-order terms. 
Then all diagrams contain only Hartree quark propagators and, hence, 
the intermediate RPA pions are massless in the chiral limit. 
As shown, e.g., in Refs.~\cite{BKRSV95,OBW00,Oer00,BBRV08} this
perturbative treatment yields the correct results for the low-temperature 
behavior of the quark condensate and the pressure.
However, since the focus of the present paper is on the phase transition,
which cannot be treated perturbatively, 
we stay with the self-consistent expansion scheme outlined above.
We should then be alerted to the fact that the intermediate RPA mesons do not 
obey the chiral Ward identities.

\section{Numerical results}\label{results}

In this section we present numerical solutions of the self-consistent 
gap equation at NLO and related quantities. The main complication as
compared to the leading-order problem arises from the fact that the
NLO self-energy correction (last diagram in \fig{gap_nlo}) is non-local.
Hence, unlike the Hartree self-energy, which only yields a constant 
shift in the mass, the self-energy is now energy and momentum dependent 
and consists of several terms with different Dirac structure.
Assuming a homogeneous medium with even parity, the inverse propagator
can be parameterized as
\begin{equation}
	S^{-1}(z, \vec k) \;=\; \gamma_0 z \;C(z, \vert\vec k \vert) - \vec{\gamma}\cdot\vec k\; A(z,\vert \vec k\vert)  - B(z,\vert \vec k\vert),
\label{Sinv}
\end{equation}
where $z$ is a complex energy variable.
In vacuum, as a consequence of Lorentz invariance, the dressing functions
$A$, $B$, and $C$ are only functions of $k^2 = z^2 - \vert\vec k \vert^2$,
and the functions $A$ and $C$ are equal. 
In the medium, however, where we have a preferred frame, we have three
independent functions, which depend on energy and momentum separately.

\subsection{Model parameters and computational details}

The integrals given in Sec.~\ref{model} are divergent and our 
model is only well-defined after specifying how to regularize them.
Since the NJL model is non-renormalizable, new cutoff parameters
can appear at each loop order. For instance, even if we have regularized
the the quark loops in the Hartree self-energy, \eq{SigmaH},
and the RPA polarization loop, \eq{Jmes},
the loop over the meson momentum in $\Sigma^{(1)}$, \eq{self_nlo},
is in general still divergent and needs to be regularized separately. 

In the following we regularize both, quark and meson loops, by sharp
3-momentum cutoffs $\Lambda$ and $\Lambda_M$, respectively. 
This has the advantage that it keeps the numerical effort
for the involved self-consistency problem as simple as possible.
Moreover, it preserves the analytic structure in the complex energy plane. 
The obvious disadvantage is that the 3-momentum cutoffs violate the Lorentz 
covariance. For the moment, we take this as a minor problem, which 
could be improved on in future modifications of the model.

The non-covariance of the regularization also makes it necessary to
specify how external 3-momenta are distributed to the propagators in
a loop. 
In the RPA polarization functions, \eq{Jmes}, we distribute $\vec q$ equally 
to both quark propagators, whereas in the NLO meson loops, \eq{self_nlo}, 
we must attribute the entire external 3-momentum to the quark propagator 
in order to be consistent with the derivation of this diagram from the 
$\Phi$-functional. 

In addition to the cutoffs, the model has two more parameters,
namely the coupling constant and the bare quark mass. We take 
\begin{equation}
	\Lambda \;=\; 664.3 \mbox{ MeV}, ~~~ G\Lambda^2 \;=\; 2.06, 
   ~~~ m_0 \;=\; 5.0 \mbox{ MeV}\,,
\end{equation}
which in leading order (Hartree / RPA) yield the empirical vacuum values 
for the pion mass $m_\pi = 135.0$ MeV and pion decay constant 
$f_\pi = 92.4$ as well as a quark condensate of 
$\langle\bar uu\rangle^{1/3} = -250.8$~MeV~\cite{Bub05}.
This corresponds to a constituent mass $m_H = 300$~MeV.
In principle, a re-fit of these parameters should be done at NLO.
However, for the mostly explorative studies of the present paper,
we keep them unchanged.
The meson loop cutoff is set to $\Lambda_M = 500.0$ MeV.

The gap equation is solved iteratively starting with a Hartree-like ansatz 
for the quark propagator. The dressing functions are stored on a grid. 
In energy direction the grid is fixed through the Matsubara frequencies and 
in 3-momentum direction an equidistant grid space of 50 MeV is chosen. 
Values in between the grid points are interpolated with cubic splines. 
The inverse propagators of the intermediate mesons are also stored on a grid.

\subsection{Dressing functions and intermediate meson propagators}
\label{dressing}

\begin{figure}
	\centering
		\includegraphics[width=0.45\textwidth]{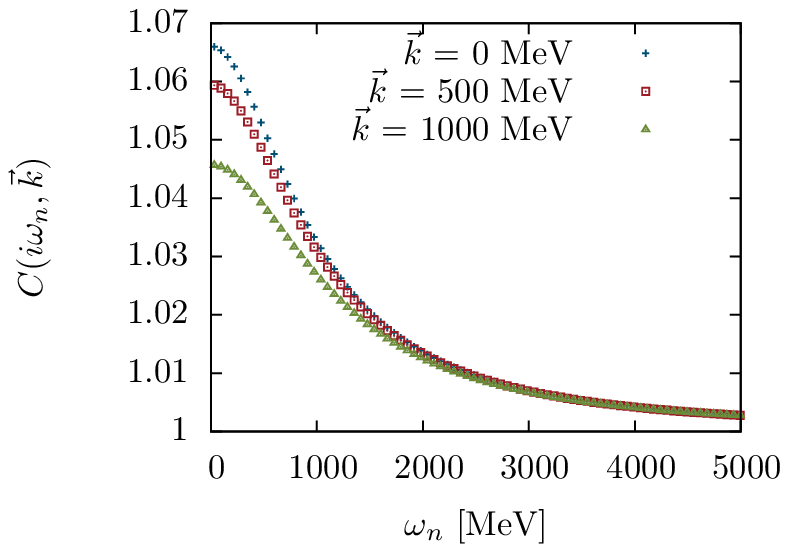}
		\includegraphics[width=0.45\textwidth]{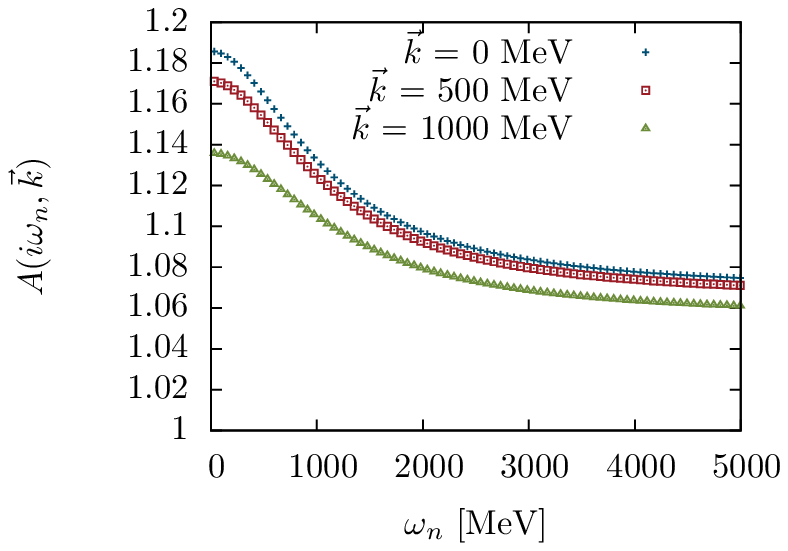}
		\includegraphics[width=0.45\textwidth]{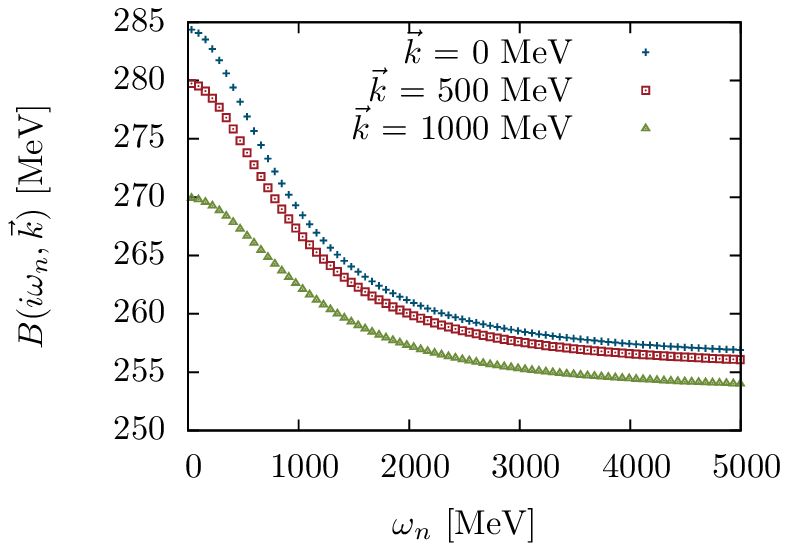}
	\caption{Dressing functions C, A, and B of the quark propagator 
at $T=10$ MeV and $\mu=$ 0 as functions of the Matsubara frequency 
for different 3-momenta.}\label{coeffnlo}
\end{figure}

Results for the dressing functions $A$, $B$, and $C$ are displayed in 
\fig{coeffnlo}. To a good approximation they can be taken to represent 
the dressing functions in vacuum, although for numerical reasons they 
have been calculated at a temperature of 10~MeV and are therefore only
given at discrete Matsubara frequencies. 
The results should be compared with the Hartree results, which are
$B_H = m_H = 300$~MeV for the present parameters and $A_H=C_H=1$.
As one can see, the NLO corrections lead to an overall reduction of 
the $B$ function, whereas $A$ and $C$ are slightly enhanced.
All dressing functions have in common that they are maximal at the lowest
$|\omega_n|$ and monotonously decrease with increasing $|\omega_n|$.  
(Note that the functions are symmetric in $\omega_n$.)
The same behavior can be observed with increasing 3-momentum.
Asymptotically, the NLO self-energy contribution vanishes, and 
the $B$ function approaches a constant value coming from the
Hartree diagram. For the same reason the $C$ function converges
to the trivial value of 1.

In principle, we would expect the same behavior for the $A$ function.
In fact, because of Lorentz covariance, the functions $A$ and $C$
should be equal in vacuum. However, as a consequence of the 
non-covariant regularization, this turns out not to be the case.
Moreover, when we consider the non-trivial parts $A-1$ and $C-1$, 
the symmetry violation is of the same order of magnitude as the physical 
effect. 
A closer inspection reveals that the $A$ function is most strongly 
affected by the cutoff artifacts, because it is directly related to the 
3-momentum (see \eq{Sinv}). In particular, this explains the wrong
asymptotic behavior of $A$. Hence, if we are interested in results which are 
sensitive to $A-1$, an improved regularization scheme should be employed.
For the quark condensate, which we discuss in Sec.~\ref{quark_condensate},
we expect that the situation is less problematic, as it is mainly
influenced by the scalar function $B$ (see \eqs{qbarq} and (\ref{TrS}) below).

\begin{figure}
	\centering
		\includegraphics[width=0.45\textwidth]{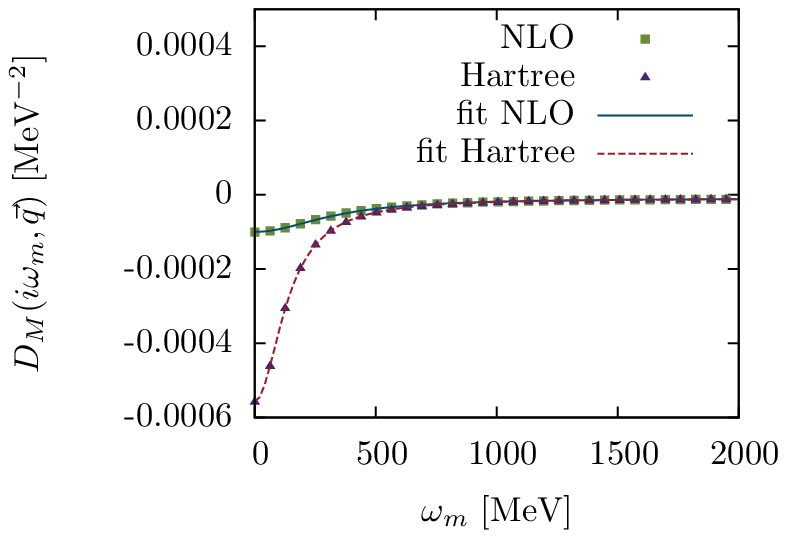}
		\includegraphics[width=0.45\textwidth]{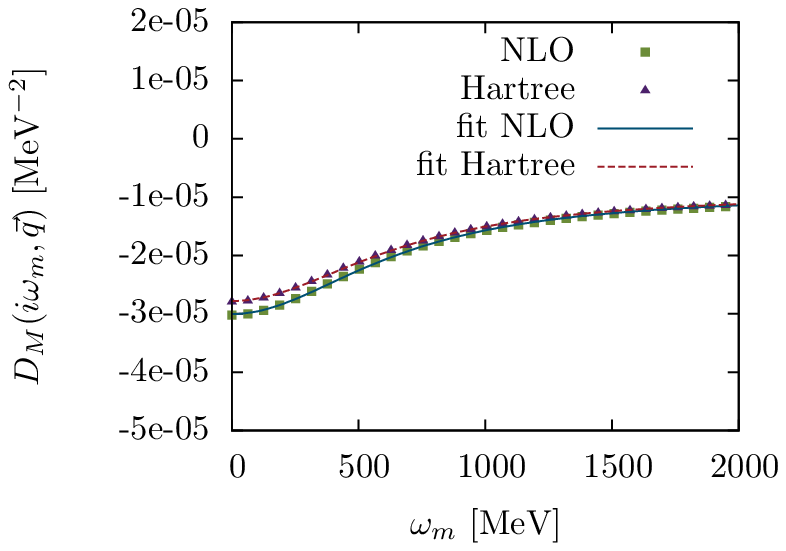}	
	\caption{Pion (upper panel) and sigma (lower panel) propagators
at $T=10$~MeV and $\mu=$ 0 as functions of the Matsubara frequency
for vanishing 3-momentum.
The intermediate RPA propagators based on the self-consistent solutions of the 
NLO gap equation are compared with the propagators in the Hartree $+$ RPA
scheme.
The numerical results at the discrete Matsubara frequencies are indicated
by points whereas the lines correspond to fits according to \eq{polefit}.}
\label{pisig}
\end{figure}

In \fig{pisig} the intermediate RPA propagators in the pion (upper panel)
and sigma (lower panel) channel channel at $T=10$~MeV and $\mu=$ 0
are displayed as functions of the Matsubara frequency for vanishing
3-momentum.
The numerical results are indicated by points.
For comparison we also show the corresponding propagators in the 
Hartree $+$ RPA scheme. As discussed at the end of Sec.~\ref{Phi} the latter
are constrained by chiral Ward identities whereas the intermediate RPA 
propagators in the NLO scheme are not. In \fig{pisig} this is reflected
by the fact that the peak of the pion propagator at $\omega_m = 0$ is
strongly suppressed in the NLO scheme: In Hartree $+$ RPA the peak is due
to the relatively near-by pole at real energies, $q^0 = m_\pi = 135$~MeV.
The strong reduction of this peak in NLO is thus a hint for a considerably 
larger mass of the intermediate pion.

We can estimate the meson masses by fitting the numerical points
with a simple pole ansatz,
\begin{equation}
D_M(i\omega_M,\vec 0) \;\approx\; 
-\frac{Z_M}{\omega_m^2 + m_M} \,-\, 2G\,.
\label{polefit}
\end{equation}
The constant $-2G$ has to be taken into account to get the correct asymptotic 
behavior (cf.~\eq{subdisp} below). 
These fits are indicated by the lines in \fig{pisig}. 
In Hartree $+$ RPA we find $m_\pi \approx 137$~MeV, in good agreement with the
true pole mass of 135~MeV. 
For the intermediate RPA pion in NLO, on the other hand, the fit yields
$m_\pi \approx 340$~MeV.

In the sigma channel, the situation is less dramatic because the sigma
meson is not a Goldstone boson. 
Here the pole fit yields $m_\sigma \approx 670$~MeV in Hartree $+$ RPA
and a slightly lower mass for the intermediate RPA sigma in NLO.
However, these numbers should not be trusted too much, as they are based on
rather far extrapolations from imaginary to real energies. (In fact, in the
MEM analysis in the next section, we find masses which are 10 -- 15\% lower.)

\begin{figure}
	\centering
		\includegraphics[width=0.45\textwidth]{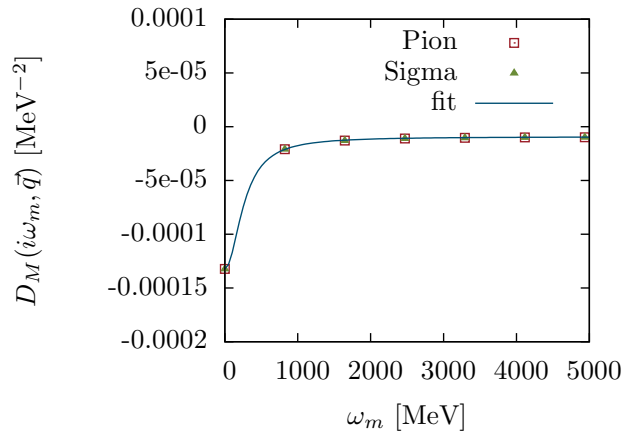}	
	\caption{Intermediate sigma and pion propagators in the NLO scheme 
in the chiral limit at $\mu=$~0 and $T=T_c=131$~MeV as functions of the 
Matsubara frequency for vanishing 3-momentum.}
\label{pisigtc}
\end{figure}

In \fig{pisigtc} we show the behavior of the intermediate sigma and pion
propagators in the chiral limit at the chiral restoration temperature,
$T=T_c$. 
As chiral symmetry is restored, sigma and pion are now degenerate.
However, even at $T_c$ the intermediate RPA mesons do not become massless.
Fitting again the numerical points with \eq{polefit},
we find $m_\pi=m_\sigma\approx 270$~MeV.
This will be relevant for the discussion below.

\subsection{The quark condensate}
\label{quark_condensate}

The chiral quark condensate
is a scalar quantity and can be calculated directly in Euclidean 
(Matsubara) space, 
\begin{equation}
	\langle \bar{q}q\rangle_T \;=\;
        T\sum_n\int\frac{d^3k}{(2\pi)^3}\mbox{Tr}(S(i\omega_n, \vec k))\,,
\label{qbarq}
\end{equation}
where 
\begin{alignat}{1}
	&\mbox{Tr}(S(i\omega_n, \vec k)) \;=\;
\nonumber\\
        &-8N_c \frac{B(i\omega_n, \vec k)}{\omega_n^2 C^2(i\omega_n, \vec k)
        + \vec k\,^2 A^2(i\omega_n, \vec k) + B^2(i\omega_n, \vec k)}
\,,
\label{TrS}
\end{alignat}
cf.~\eq{Sinv}.

\begin{figure}
	\centering
		\includegraphics[width=0.45\textwidth]{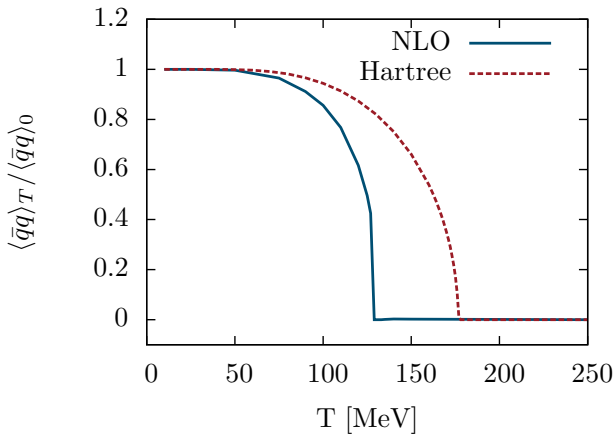}	
	\caption{Temperature dependence of the chiral quark
condensate in the chiral limit in Hartree approximation (dashed) and in NLO
(solid) at $\mu=0$.}
\label{condT}
\end{figure}

Our results for its temperature dependence in the chiral limit are displayed 
in \fig{condT}, both, in Hartree approximation (dashed) and in NLO (solid). 
We find that chiral symmetry is restored in a second-order phase transition 
in both cases. The critical temperature is decreased by the NLO corrections.

A second-order phase transition is also what is expected for two-flavor QCD
in the chiral limit. At the critical temperature the fermionic degrees of 
freedom are suppressed due to their antiperiodicity. 
Therefore the phase transition is dominated by four bosonic degrees of freedom 
(three pions and the sigma meson) which are all massless at $T_c$
and one expects critical behavior according to the $O(4)$ universality 
class~\cite{PW84}.  
Since these arguments are only based on the symmetries and dimensionality
of the system, the same should hold in any theory or model with the same 
conditions. 
Indeed, a second-oder phase transition with $O(4)$ critical exponents
has been found in a renormalization group (RG) approach to the two-flavor
quark meson model~\cite{Schaefer:1999em, Schaefer:2006ds}, and the NJL model should 
in principle exhibit a similar behavior. 

However, it is not a priori clear to what extent the universal
behavior is spoiled by the truncation scheme. 
As we have seen, the massless bosonic degrees of freedom which are the basis
of the universality arguments are not manifest in the gap equation, 
neither in Hartree approximation nor in NLO:
In the Hartree self-energy, there is no back-reaction of the RPA mesons on 
the quark propagator, whereas the intermediate RPA mesons which enter the NLO 
gap equation are not massless. 
From this point of view, it seems not even guaranteed that the phase 
transition must be second order in these approximation schemes. 
To understand why this nevertheless should be expected, we can adopt
the arguments of Ref.~\cite{ABC04}, where the chiral phase transition was
investigated in a purely bosonic model at NLO and found to be second order
as well: As we have discussed, at each order one can in
principle construct mesonic correlators with the correct chiral behavior 
by applying \eq{phik} to the $\Phi$-functional and iterating the resulting
scattering kernel in the BSE. One could then employ these correlators to
study critical exponents. Obviously, this can only work if the phase 
transition is second order.  Thus, the gap equation must somehow ``know'' 
about the massless degrees of freedom even if they do not enter the equation
explicitly.

In this context the consistency of the approximation scheme is crucial:
In Refs.~\cite{FB96,Oer00} a first-order phase transition was found in a 
simplified NLO scheme, which was suggested in Refs.~\cite{DSTL95,NBC+96}.
In that scheme, only local contributions to the quark self-energy are 
taken into account. 
Although one can formally show that there are massless Goldstone bosons in
the chiral limit, the approximation is not thermodynamically consistent
as the gap equation cannot be derived from a thermodynamic potential.
This suggests that a thermodynamically consistent treatment is important
to find the correct order of the phase transition.

Yet, even in a self-consistent and thermodynamically consistent
truncation scheme, not all details of the critical behavior
are necessarily reproduced correctly.
It is well known that the Hartree gap equation, although correctly 
predicting a second-order phase transition, yields mean-field critical
exponents.
At NLO, there might be some improvement, but we should not expect
to find exact $O(4)$ behavior. 
Unfortunately, our numerical results are not precise enough to work this
out quantitatively.

The low-temperature behavior of the quark condensate is model independently
given by chiral perturbation theory. For two quark flavors one finds
in the chiral limit~\cite{GL87}
\begin{equation}
\langle \bar q q\rangle_T \;=\; 
\langle \bar q q\rangle_0 \left(1-\frac{T^2}{8f_\pi^2} 
- \frac{T^4}{384 f_\pi^4} + ...\right)\,,
\label{qbqqXPT}
\end{equation}
where $\langle \bar q q\rangle_0$ is the condensate at $T=0$ and 
$f_\pi$ is the pion decay constant.
This behavior is entirely due to the massless chiral pions. 
The $T^2$ term corresponds to their ideal gas contribution,
while the $T^4$ term is due to $p$-wave $\pi-\pi$ interactions.

Obviously, this behavior cannot be reproduced in the Hartree approximation,
which includes no back-coupling of the mesons to the quark propagator.
The change of the condensate is then exclusively triggered by thermal 
quarks and, hence, exponentially suppressed due to their mass. 

On the other hand, it was shown in Ref.~\cite{Oer00} that at least the
$T^2$ term in \eq{qbqqXPT} is reproduced correctly, if the $1/N_c$ 
corrections to the quark propagator are taken into account perturbatively. 
To be more precise, it was shown that the leading correction corresponds
to that of an ideal gas of RPA pions. 
Since in the perturbative approach the latter are built from Hartree
quarks, they are massless in the chiral limit and, thus, lead to the
correct low-temperature behavior.

Accordingly, the $T^2$-term in \eq{qbqqXPT} cannot be reproduced in
our fully self-consistent NLO scheme. As in the perturbative approach,
there are corrections from the intermediate RPA mesons. However,
because of their relatively large masses, their effect is exponentially
suppressed.  
We are thus faced with the situation that the self-consistent scheme 
gives only a poor description of the low-temperature behavior but 
works well at the phase transition, 
while it is just the other way around in the perturbative approach.

\begin{figure}
	\centering
		\includegraphics[width=0.45\textwidth]{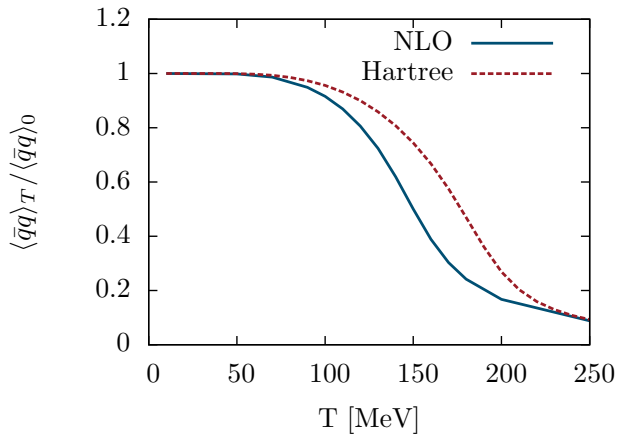}	
	\caption{Temperature dependence of the quark condensate for a 
bare quark mass $m_0 = 5$~MeV in Hartree approximation (dashed) and 
NLO (solid) at $\mu = 0$.}
\label{condf}
\end{figure}

In \fig{condf} we show the temperature dependence of the quark condensate
for the case of a non-vanishing bare quark mass. 
In this case, chiral symmetry is only approximately restored in a 
crossover. Again, the NLO corrections lead to a reduction of the
cross-over temperature relative to the Hartree result.

\begin{figure}
	\centering
		\includegraphics[width=0.45\textwidth]{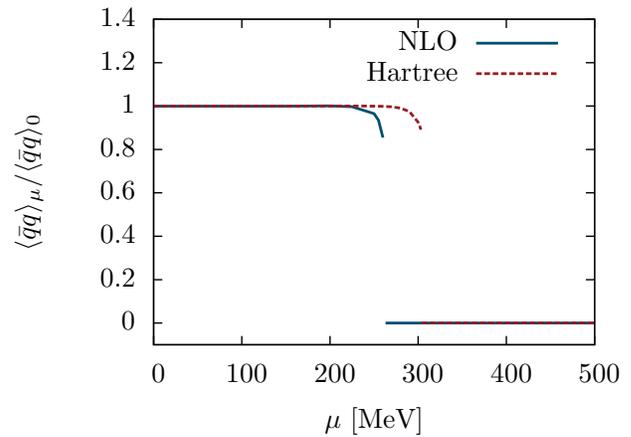}	
	\caption{Quark condensate in the chiral limit as a function of the 
         chemical potential in Hartree approximation (dashed) and NLO
         (solid) at $T=0$.}
\label{condmu}
\end{figure}

The dependence of the quark condensate on the chemical potential at $T=0$
can be seen in \fig{condmu} for the chiral limit. 
The Hartree and NLO results look qualitatively similar. In both cases chiral 
symmetry is restored in a first-order phase transition. Similar to the 
temperature behavior, the critical chemical potential is slightly lower in 
NLO. 

\section{Analytic continuation with the Maximum-Entropy-Method (MEM)}\label{MEM}

Searching the analytic continuation of a propagator given at a discrete set of Matsubara frequencies is an ill-posed problem. In general, an infinite set of functions would provide an analytic continuation. Additional requirements on asymptotics and analytical structure provide a unique solution \cite{BM61} but this only helps if one can get analytic expressions for the functions.

One way to attack this problem numerically is the Maximum-Entropy Method (MEM).
The quantity to calculate is the spectral function $\rho(\omega)$ which is 
related to the propagator via the Lehmann representation (e.g. \cite{Bel00}),
\begin{equation}\label{leh}
D(z)\;=\;\int\limits_{-\infty}^{\infty}\frac{d\omega}{2\pi}
\frac{\rho(\omega)}{z-\omega}\,,
\end{equation}
and can be interpreted as a probability distribution. The method performs a 
$\chi^2$-fit to the discrete ``data'' $D(i\omega_n)$
and additionally requires minimum deviation from the so-called prior estimate,
the ``most probable'' spectral function in absence of any data.
The combination of both requirements finally leads to a unique solution.
MEM has been successfully used in lattice QCD \cite{AHN01, WK+02} and Dyson-Schwinger calculations \cite{Nic07}.

\subsection{The Method}

We largely follow the formalism described in Refs.~\cite{AHN01,Nic07}.
The basis for MEM is Bayes' theorem for conditioned probability applied to the plausibility $P[\rho|D H(m)]$ of a spectral function $\rho$ under given data $D$ and a prior knowledge $H(m)$,
\begin{equation}\label{bayes}
	P[\rho|D H(m)] \;=\; \frac{P[D|\rho H(m)]P[\rho|H(m)]}{P[D|H(m)]}\,.
\end{equation}
$P[D|\rho H(m)]$ is the likelihood function, which indicates the plausibility of the data under the spectral function and the prior and $P[\rho|H(m)]$ is the prior probability for the plausibility of $\rho$ solely under the prior. $P[D|H(m)]$ is a normalization factor independent of the spectral function and can be dropped as the probabilities are normalized in the end. 

For the likelihood function a Gaussian distribution is assumed
\begin{equation}\label{likelihood}
	P[D|\rho H(m)] \;=\; \frac{1}{Z_L}e^{-L[\rho]}
\end{equation}
with
\begin{equation}\label{Lmem}
	L[\rho] \;=\; \frac{1}{N_{D}}\sum_i \frac{\vert D_i - D^{\rho}_i \vert^2}{2\vert\sigma_i\vert^2}
\end{equation}
for $N_D$ equidistant ``data'' with values $D_i$ and errors $\sigma_i$ and the
corresponding values $D^\rho_i$ calculated from the given spectral function 
$\rho$ using the Lehmann representation, \eq{leh}.
$Z_L$ is a normalization constant. Note that, by this assumption, 
$P[D|\rho H(m)]$ does actually not depend on on the prior. 
 
The prior probability $P[\rho|H(m)]$ depends directly on a prior estimate 
$m(\omega)$ for the spectral function. $m(\omega)$ contains general information about spectral functions, especially positivity, and is usually chosen as a constant function. With help of a scale factor $\alpha$, $P[\rho|H(m)]$ can be expressed as
\begin{equation}
P[\rho|H(m)] \;=\; 
\int \limits_0^\infty d\alpha P[\rho|H(\alpha m)] P[\alpha|H(m)]\,,
\end{equation}
where
\begin{equation}\label{priorprob}
	P[\rho|H(\alpha m)] \;=\; \frac{1}{Z_S} e^{\alpha S[\rho]}
\end{equation}
with a normalization constant $Z_S$ and the Shannon-Jaynes entropy
\begin{equation}
	S[\rho] \;=\; \int \limits_{-\infty}^\infty d\omega \left( \rho(\omega) - m(\omega) -\rho(\omega) \ln{\frac{\rho(\omega)}{m(\omega)}} \right)\,.
\end{equation}
This can be derived axiomatically by using general features of the entropy
(locality, scale invariance etc.) or with help of the law of large numbers
(``monkey argument'') \cite{AHN01}. Discretization and expanding the logarithm 
for small deviations of the spectral function from the prior yields
\begin{equation}
	S[\rho] \approx -2\sum_i \Delta\omega_i \left( \sqrt{\rho_i} - \sqrt{m_i}\right)^2\,.
\end{equation}
Applying \eq{likelihood} and \eq{priorprob} to \eq{bayes} gives
\begin{equation}\label{Qmem}
	P[\rho|D H(\alpha m)] \;=\; \frac{1}{Z} e^{Q[\rho]}
\end{equation}
with another normalization constant $Z$ and
the functional $Q[\rho] = \alpha S[\rho] - L[\rho]$. 
Maximizing $Q[\rho]$ gives the most probable spectral function, 
$\rho(\omega) = \rho_\alpha(\omega)$, for given $\alpha$.

Finally the scale factor $\alpha$ has to be eliminated, which can be done in 
several ways. We use Bryan's method \cite{Bry90}, which is applied in most 
cases. Here, the final spectral function $\rho_{MEM}(\omega)$ is obtained by 
averaging over $\alpha$,
\begin{alignat}{1}
	\rho_{MEM}(\omega) &\;=\; \int \mathit{D}\rho\, \rho(\omega) P[\rho|D H(m)]
\nonumber\\
	&\approx \int\limits_0^\infty d\alpha\, \rho_\alpha(\omega) P[\alpha|D H(m)]\,,
\end{alignat}
with the probability factor
\begin{alignat}{1}
\label{probmem}
	P[\alpha|D H(m)] & \propto \int \mathit{D}\rho\, e^{Q[\rho]} 
\nonumber\\
	& \approx \exp\left[\frac{1}{2}\sum\limits_k\ln\left(\frac{\alpha\Delta\omega_k}{\lambda_k}\right) + Q[\rho_\alpha] \right]\,.
\end{alignat}
and $\lambda_k$ being the eigenvalues of the matrix
\begin{equation}
	M_{ij} \;=\; \alpha\Delta\omega_i\delta_{ij} + \sqrt{\rho_i} \left. \frac{\partial^2 L}{\partial\rho_i\partial\rho_j} \sqrt{\rho_j}\right\vert_{\rho = \rho_\alpha}
\end{equation}
In an intermediate step Laplace's rule ($P[\alpha|H(m)]= const.$) \cite{AHN01} has been applied. For calculating $\rho_{MEM}(\omega)$ the probabilities \eq{probmem} have to be normalized.

\subsection{Numerical implementation}

The iterative calculation of the quark propagator discussed in 
Sec.~\ref{results}
has been performed without error estimate, as the largest parts of the error
are highly correlated and systematical. MEM requires data with noncorrelated
Gaussian errors (\eq{Lmem}). Therefore we assume a constant relative error of
$10^{-4}$ all data points. This error underestimates the systematical errors
but should be of the order the uncorrelated numerical errors. This choice
seems useful as larger errors do not resolve the continuum contributions to
the spectral functions while smaller errors lead to unphysical oscillations.

We chose a constant prior of $m(\omega) = 10^{-3}$~MeV$^{-1}$. This arbitrary choice can be justified as the result is quite insensitive, even when the
prior is varied by six orders of magnitude, as illustrated in \fig{specprior}.

\subsection{Vacuum spectral functions}
\label{MEM_vac}

\begin{figure}
	\centering
		\includegraphics[width=0.45\textwidth]{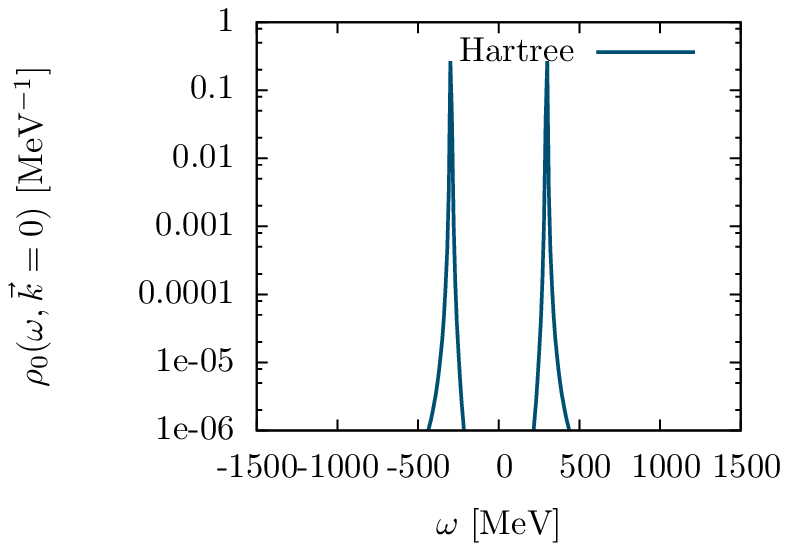}	
		\includegraphics[width=0.45\textwidth]{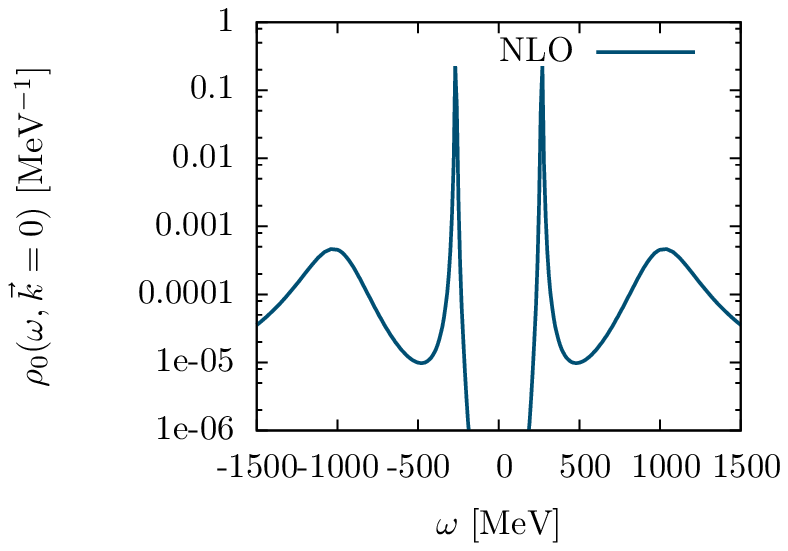}	
	\caption{MEM results for the quark spectral function $\rho_0(\omega)$
in Hartree approximation (upper panel) and NLO (lower panel) in vacuum
for vanishing 3-momentum.}
\label{spec0vac}
\end{figure}

Vacuum\footnote{As before, the ``vacuum'' results have been obtained at 
$\mu=0$ and a temperature $T=10$~MeV for numerical reasons. We expect
the difference to real vacuum calculations at $T=0$ to be small.}
results for the quark spectral function are shown in \fig{spec0vac}.
To be precise, since the spectral function of spin-$\frac{1}{2}$ fermions
has a Dirac structure, we show the 0-component,
\begin{equation}
\rho_0 \;=\; \frac{1}{4}\mathrm{Tr}(\gamma^0\rho)\,.
\end{equation}
Moreover, we consider quarks with vanishing 3-momentum.
 
The Hartree spectral function is displayed in the upper panel. Here we can
directly compare the MEM result with the exact analytical solution,
\begin{equation}
\rho_0(\omega) = \pi \delta(|\omega|-m_H)\,,
\label{specH}
\end{equation}
i.e., two delta peaks at $\omega = \pm 300$~MeV. 
Indeed, MEM yields two sharp peaks at the correct positions, although with
a small width which is caused by the assumed numerical errors.

\begin{figure}
	\centering
		\includegraphics[width=0.45\textwidth]{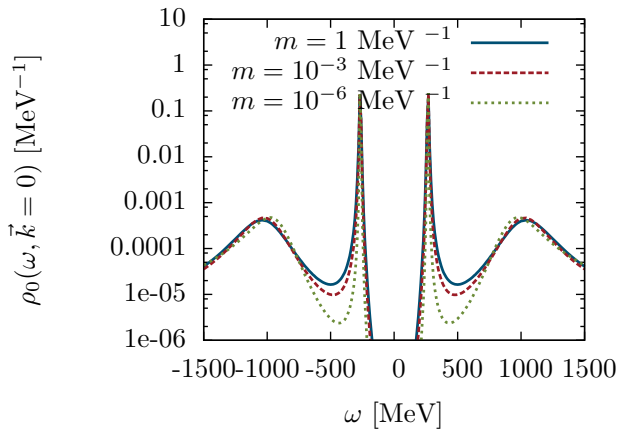}		
	\caption{MEM result for the NLO quark spectral function 
$\rho_0(\omega)$ in vacuum for vanishing 3-momentum, using different priors 
$m(\omega)=\mathit{const}$.}
\label{specprior}
\end{figure}

The NLO spectral function (lower panel) has similar peaks, shifted to 
slightly lower energies, $\omega \approx \pm 270$~MeV.
In addition there are two broad bumps at higher values of $|\omega|$.
These structures can be related to the imaginary part of the 
NLO self-energy diagram (last diagram of \fig{gap_nlo}), i.e.,
to the continuum due to meson absorption or emission processes on the 
quark.
This continuum should be well separated from the quark mass peak and start
at a threshold given by the sum of the quark mass and the intermediate pion 
mass. According to our earlier estimate for $m_\pi$, this would be at
around 600~MeV and is more or less consistent with the MEM result.
The details of the threshold region can, however, not be resolved. 
From \fig{specprior} we also see that the dip region, where we expect a 
vanishing spectral function, is most sensitive to the prior.

Unlike in Hartree approximation, the MEM results for the spectral functions
in the self-consistent NLO scheme cannot be confronted with direct 
calculations in Minkowski space, since the latter are highly non-trivial. 
However, such a comparison can be done for the quark propagator with 
{\it perturbative} NLO corrections to the self-energy.
In that case, as already mentioned, only the Hartree propagator $S_H$ is
calculated self-consistently and afterwards perturbative corrections 
are added to the self-energy, 
\begin{equation}
	S^{-1}_{pert} \;=\; S^{-1}_{H} - \Sigma^{(1)}_{pert}\,,
\end{equation}
where $\Sigma^{(1)}_{pert}$ is strictly of the order $1/N_c$.
In particular, it is entirely given in terms of $S_H$ 
and meson propagators in the Hartree $+$ RPA scheme.
This makes its evaluation in Minkowski space possible.

As discussed, e.g., in Ref.~\cite{Oer00}, $\Sigma^{(1)}_{pert}$ consists 
of two diagrams. The first one is local and, thus, only gives a constant 
real contribution to the mass. In the following comparison we will 
therefore neglect this term for simplicity.
The second contribution is non-local and given by self-energy in the last 
diagram in \fig{gap_nlo} if all self-consistent quark propagators are replaced 
by $S_H$.   
In Minkowski space we make the additional approximation of neglecting the 
continuum parts and finite widths of the mass peaks of the RPA mesons.
As the mass peaks are the dominant contributions to the spectrum 
and even for the unstable sigma meson the main peak is very sharp, 
this approximation should be quite accurate. 

\begin{figure}
	\centering
		\includegraphics[width=0.45\textwidth]{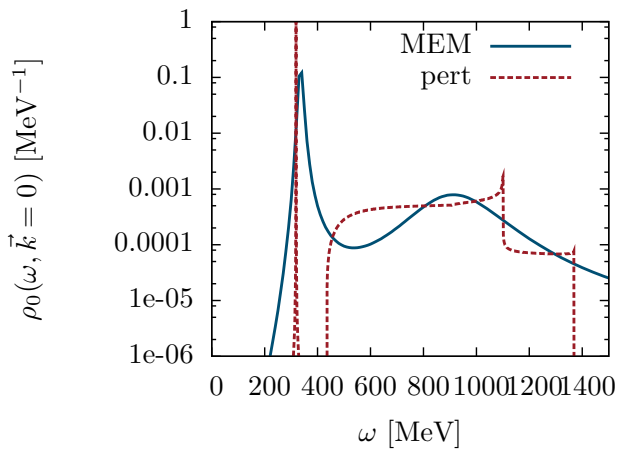}	
	\caption{Comparison between MEM result and direct Minkowski-space
calculation of the quark spectral function with perturbative $1/N_c$
corrections to the self-energy in vacuum.}
\label{mempert}
\end{figure}

A comparison with the MEM output of a corresponding perturbatively dressed 
propagator calculated in Euclidean space is shown in \fig{mempert}. For 
numerical reasons the calculation in Minkowski space has been performed with 
quark propagators with a width of 1~MeV, which was added by hand. 
The main peak of the MEM result and the 
Minkowski result are almost at the same energy. (The small difference could
be due to the mentioned approximations for the meson spectral function 
in Minkowski space.) The continuum contributions only show rough agreement 
in height and position but the shape differs considerably. The main reason for 
this is the dominance of the mass peak in the spectrum. The contina give only 
small contributions and it would require data with much lower errors to 
become sensitive to the shape of the continuum.
Furthermore the continuum structure is very complicated with sharp thresholds 
and peaks which are difficult to be reproduced by MEM.
In fact, the high-energy thresholds and peaks are artifacts of the 
regularization and therefore not even physical.
In principle, one could try to include these effects into the prior
$m(\omega)$ as done in the lattice calculations of Ref.~\cite{Ding:2009ie}.

For applying MEM to meson propagators we use a subtracted dispersion relation,
\begin{equation}
	D_M(z)\;=\;\int\limits_{-\infty}^{\infty}\frac{d\omega}{2\pi}\frac{\rho(\omega)}{z-\omega} - 2G\,,
\label{subdisp}
\end{equation}
which takes care of the asymptotic behavior of NJL RPA meson propagators. 

\begin{figure}
	\centering
		\includegraphics[width=0.45\textwidth]{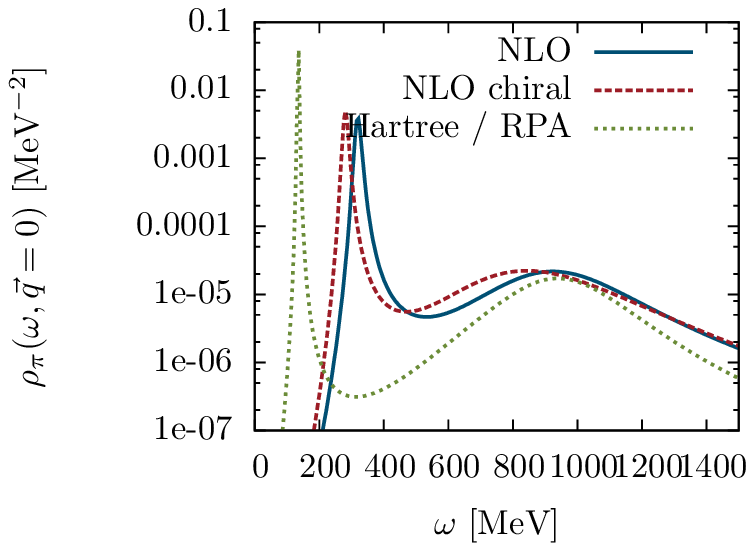}	
		\includegraphics[width=0.45\textwidth]{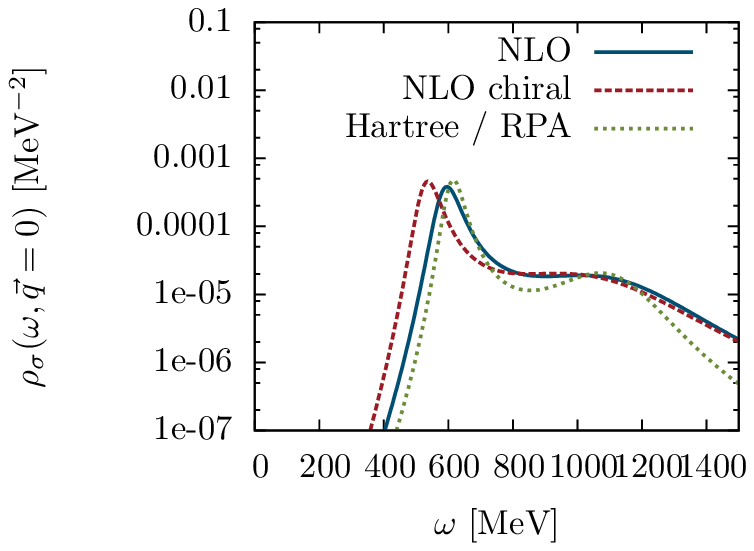}	
	\caption{MEM results for the RPA pion (upper panel) and sigma 
(lower panel) spectral function in vacuum for vanishing 3-momentum.
The various curves correspond to the intermediate RPA mesons in the NLO
scheme for $m_0 = 5$~MeV and in the chiral limit, as well as in the 
Hartree $+$ RPA scheme for $m_0 = 5$~MeV.}
\label{specps}
\end{figure}

The spectral functions for the RPA meson propagators are displayed 
in \fig{specps}. The intermediate RPA mesons in the NLO scheme 
with finite current quark mass and in the chiral limit are
compared with the mesons in the Hartree $+$ RPA scheme. 

The latter can also be calculated directly in Minkowski space
(cf. Sec.~\ref{MEM_med}). 
In the pion channel this yields a delta peak at 
$\omega = m_\pi = 135$~MeV and a continuum due to quark-antiquark
decay above a threshold of $2m_H = 600$~MeV.
As one can see in the figure (upper panel, dotted line), these
features are again qualitatively reproduced by MEM: 
We find a sharp peak at 135~MeV and a broad bump at higher energies,
which can be identified with the continuum. On the other hand, as
before, the detailed threshold behavior cannot be reproduced. 
 
The spectral function of the intermediate RPA pion in NLO (solid line)
looks qualitatively similar, but the mass peak is now located at 
$\omega = 320$~MeV. This is slightly lower than our estimate by the
pole fit in Sec.~\ref{dressing} (340~MeV), but still very heavy. 
As we have said repeatedly, this reflects the fact that the intermediate RPA
mesons are not constrained by chiral Ward identities.
In fact, even in the chiral limit (dashed line), we find a mass peak
at 280~MeV in the MEM spectral function.  

In the sigma channel, the exact spectral function in Hartree $+$ RPA
has a mass peak slightly above the continuum threshold. This is again
qualitatively reproduced by MEM (lower panel, dotted line), where we find 
a maximum at about 615~MeV. 
The NLO corrections to the sigma are rather small (solid line),
the mass is lowered by about 20~MeV. 
In the chiral limit (dashed line), it is further reduced by
60~MeV, but the gross features of the spectral function remain unchanged.

\subsection{In-medium spectral functions}
\label{MEM_med}

Larger temperatures increase the inaccuracies significantly as the larger 
distance between the Matsubara frequencies ($2\pi T$) provides only few data 
at low frequencies while high-frequency data mainly carry information about 
the asymptotics. 

\begin{figure}
	\centering
		\includegraphics[width=0.45\textwidth]{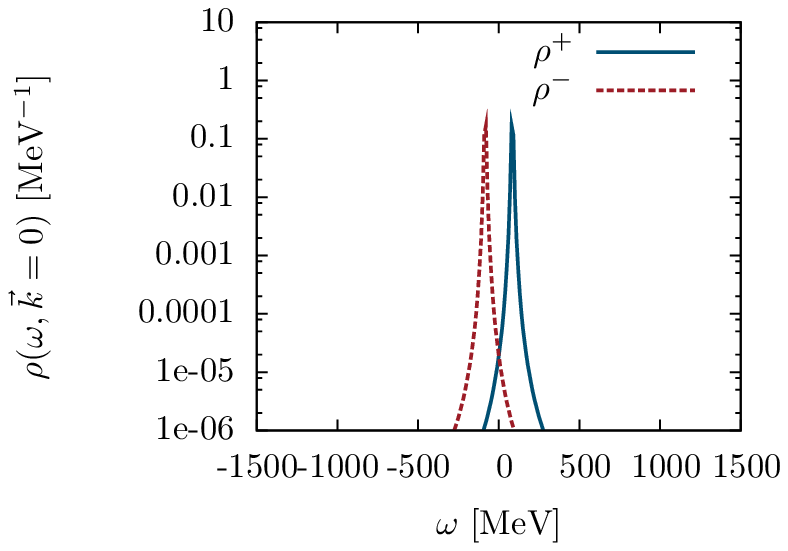}	
		\includegraphics[width=0.45\textwidth]{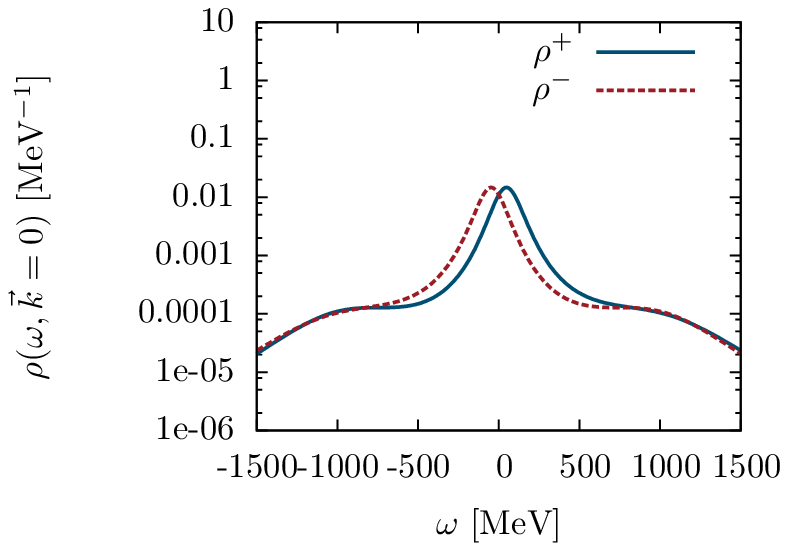}	
	\caption{MEM result for the quark spectral functions $\rho^{}(\omega)$
in Hartree approximation (upper panel) and NLO (lower panel) 
for vanishing 3-momentum at $T = 200$~MeV and $\mu = 0$. 
The solid (dashed) lines correspond to an assumed 
relative numerical error of $10^{-4}$ ($10^{-6}$).} 
\label{spec0T200}
\end{figure}

As for larger temperature the quark mass decreases and therefore the main
peaks in the spectral function come closer together, extrapolating the 0-component
of the spectral function with MEM does not resolve the single peaks. Therefore it 
is reasonable to study other projections of the spectral function to
separate the peaks. For vanishing 3-momentum this can be achieved by the
decomposition of the spectral function
\begin{equation}
	\rho(\omega) = \rho^+(\omega)L^+\gamma_0 + \rho^-(\omega)L^-\gamma_0
\end{equation}
with the projectors $L^\pm = \frac{1}{2}(1\pm\gamma_0)$. The components 
$\rho^\pm$ correspond to the spectra of particle and antiparticles and are
also positive definite, therefore MEM is applicable. In Hartree approximation
the analytic result yields
\begin{equation}
	\rho^+(\omega) = \pi\left(1+\frac{m}{\omega}\right)\delta(\omega - m)
\end{equation}
\begin{equation}
	\rho^-(\omega) = \pi\left(1-\frac{m}{\omega}\right)\delta(\omega + m)
\end{equation}
and the two main peaks are separated into the different channels. This decomposition
has also been used in recent lattice studies at finite temperature \cite{Karsch:2009tp}.

MEM results for $T = 200$ MeV are shown in \fig{spec0T200}. While in Hartree approximation
the quark and antiquark peaks are still well-separated the NLO peaks are much
broader and overlap. This is an indicator for the thermal broadening of the quark
spectral function in the medium.

\begin{figure}
	\centering
		\includegraphics[width=0.45\textwidth]{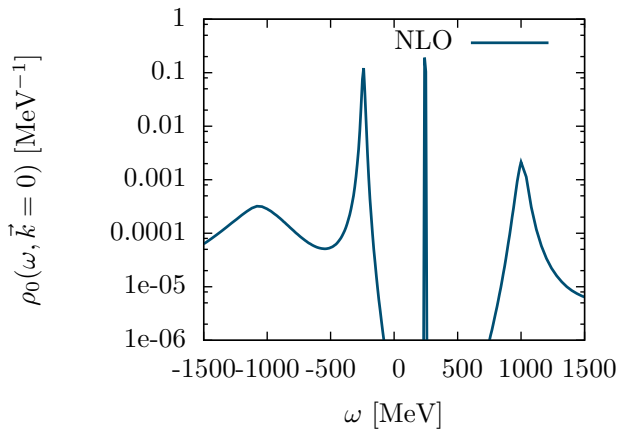}	
	\caption{MEM result for the NLO quark spectral function $\rho_0(\omega)$
for vanishing 3-momentum at $T=0$ and $\mu = 200$~MeV.}
\label{spec0mu}
\end{figure}

Applying MEM at finite chemical potential visualizes the shifted Fermi surface. 
The quark spectral function at $\mu = 200$ MeV is displayed in \fig{spec0mu}.
We obtain a quite asymmetric result: At positive energies we find a very sharp 
mass peak which is well separated from the continuum at higher energies.
At negative energies, on the other hand, both, the mass peak and the continuum
are broader and the dip region in between is less pronounced. 

Although at finite chemical potential the spectral function is indeed in
general no longer symmetric, 
the main reason for the observed asymmetry is probably caused by the
MEM procedure. This is due to the fact that the dispersion relation is now
of the form
\begin{equation}
	S(i\omega_n + \mu) \;=\; \int\limits_{-\infty}^{\infty}\frac{d\omega}{2\pi}\frac{\rho(\omega)}{i\omega_n + \mu - \omega}\,.
\end{equation}
Therefore, the integral is most sensitive to details of the spectral function
near the Fermi surface, where it gives the largest contribution.
Hence, in the present example, the MEM procedure works best around 
$\omega = 200$~MeV, which explains why the positive mass peak is much better
resolved.

\begin{figure}
	\centering
		\includegraphics[width=0.45\textwidth]{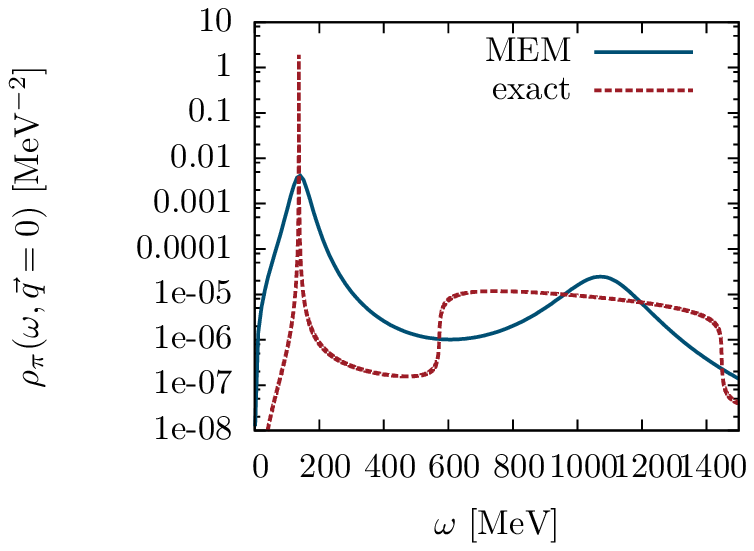}	
		\includegraphics[width=0.45\textwidth]{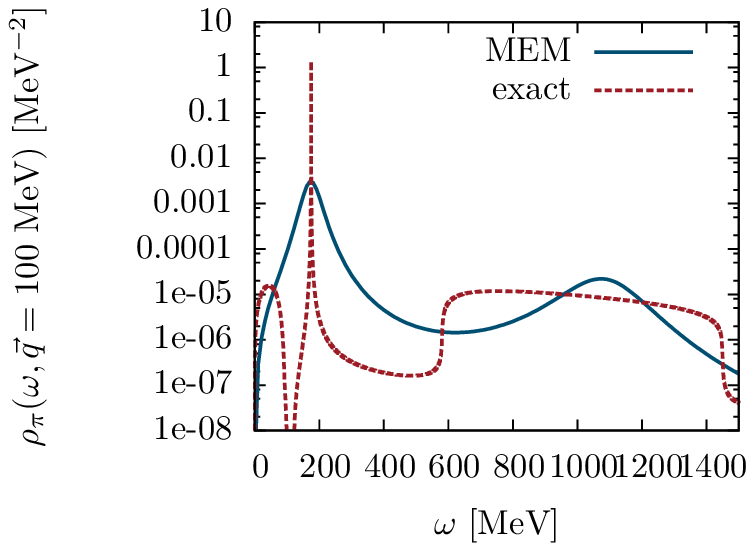}	
	\caption{Hartree $+$ RPA pion spectral function in directly calculated 
in Minkowski space and extrapolated with MEM at $T = 100$ MeV for 3-momentum 
$|\vec q| = 0$ (upper panel) and $|\vec q| = 100$~MeV (lower panel).}
\label{pimem100}
\end{figure}

In the mesonic sector, Hartree $+$ RPA meson spectral functions can be 
calculated directly in Minkowski space without further assumptions. 
A comparison with MEM results in the pion channel at $T = 100$~MeV
is displayed in \fig{pimem100} 
for $|\vec q| = 0$ (upper panel) and $100$ MeV (lower panel). 
In the Minkowski-space calculations, a width of 1~MeV was again added by 
hand to the propagators for numerical reasons.

We find that the agreement of the MEM results with the direct calculations
is rather poor. Only the position of the main peak fits while the shape of the 
spectral function looks totally different. The spacelike particle-hole branch 
in the spectrum which occurs at finite 3-momentum cannot be resolved in the 
MEM output. 
 
\begin{figure}
	\centering
		\includegraphics[width=0.45\textwidth]{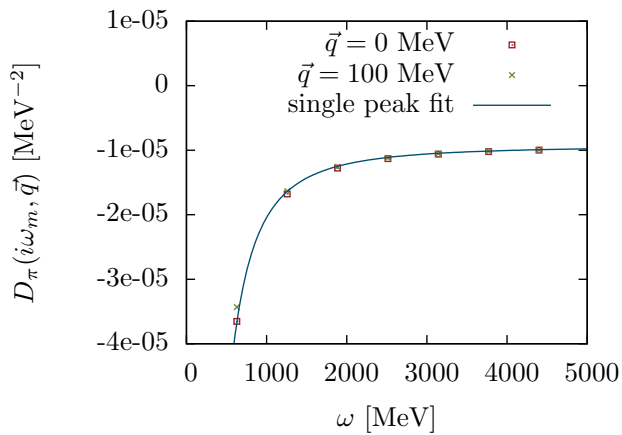}	
	\caption{Hartree $+$ RPA Pion propagator in Euclidean space at 
$T=100$~MeV. The line is the data coming from a spectral function with a single delta peak.}\label{mempertprop}
\end{figure}

In order to shed some light on this problem, we show in \fig{mempertprop}
the Euclidean data which served as input for the MEM results in 
\fig{pimem100}. 
At low Matsubara frequencies the data points with different 3-momenta are
slightly shifted against each other, which takes care of the different mass 
peak positions. 
But one cannot see a qualitative difference which could produce the spacelike 
continuum contributions which exist at  $|\vec q| = 100$~MeV, but not at
$|\vec q| = 0$. Furthermore the complicated structure of the spectral 
functions is not visible in the Euclidean data and so this structure gets 
almost lost in the convolution with the Lehmann representation \eq{leh}. 
Instead the fit of a propagator with a single mass pole according to a delta 
peak in the spectrum already fits the data almost perfectly. This exemplifies 
how difficult it is to regain the spectral function from the Euclidean data.

\section{Summary}\label{summary}
We have studied the NJL model in next-to-leading order in a self-consistent $1/N_c$-expansion. The chiral condensate showes a second-order phase transition at finite temperature and vanishing chemical potential in agreement with expectations from the O(4) universality class. In comparison to mean-field results the critical temperature is decreased. The low temperature behavior expected from chiral perturbation theory cannot be reproduced as the intermediate mesons which enter the NLO quark self-energy diagram are not restricted in terms of chiral symmetry and are very massive.
At finite chemical potential and vanishing temperature the phase transition is of first order like in mean-field approximation. The critical chemical potential is also slightly reduced. 

The Maximum-Entropy-Method has been used to calculate spectral functions for real energies from the Euclidean propagators. The mass peak of the particles can be reproduced quite well and for NLO quark propagators also a continuum contribution can be identified. The large mass of the intermediate pions is confirmed and approximately of order $\sim 320$ MeV while the masses of the quarks ($\sim 270$ MeV) and intermediate sigma mesons ($\sim 600$ MeV) are of the order of the mean-field values. Finite temperature increases the inaccuracies of MEM as fewer data at low energies are available. For perturbatively dressed propagators the spectral function can be calculated directly in Minkowski space. A comparison with these results shows that the main peak is reproduced well by MEM, but not the continua. 

\begin{acknowledgments}
We thank J\"urgen Berges, Bengt Friman, Hendrik van Hees, J\"orn Knoll and Andrey Radzhabov for usefull discussions and comments about chiral symmetry and universality and Dominik Nickel for his help with the Maximum entropy method. D.M. was supported by BMBF under contract 06DA9047I and by the Helmholtz Graduate School for Hadron and Ion Research. We also acknowledge support by the Helmholtz Alliance EMMI and the Helmholtz International Center for FAIR.

\end{acknowledgments}



\begin{thebibliography}{99}

\bibitem{NJL61}
  Y.~Nambu and G.~Jona-Lasinio,
  Phys.\ Rev.\  {\bf 122}, 345 (1961);
  Phys.\ Rev.\  {\bf 124}, 246 (1961).

\bibitem{Hands:2002mr}
  S.~Hands and D.~N.~Walters,
  Phys.\ Lett.\  B {\bf 548}, 196 (2002)
  [arXiv:hep-lat/0209140].

\bibitem{VW91}
  U.~Vogl and W.~Weise,
  Prog.\ Part.\ Nucl.\ Phys.\  {\bf 27}, 195 (1991).

\bibitem{Klev92}
  S.~P.~Klevansky,
  Rev.\ Mod.\ Phys.\  {\bf 64}, 649 (1992).

\bibitem{HK94}
  T.~Hatsuda and T.~Kunihiro,
  Phys.\ Rept.\  {\bf 247}, 221 (1994)
  [arXiv:hep-ph/9401310].

\bibitem{Iwasaki:2006dr}
  M.~Iwasaki, H.~Ohnishi and T.~Fukutome,
  arXiv:hep-ph/0606192.

\bibitem{QK94}
  E.~Quack and S.~P.~Klevansky,
  Phys.\ Rev.\  C {\bf 49}, 3283 (1994).

\bibitem{ENV94}
  D.~Ebert, M.~Nagy and M.~K.~Volkov,
  Phys.\ Atom.\ Nucl.\  {\bf 59}, 140 (1996)
  [Yad.\ Fiz.\  {\bf 59}, 149 (1996)]
  [arXiv:hep-th/9412214].

\bibitem{DSTL95}
  V.~Dmitrasinovic, H.~J.~Schulze, R.~Tegen and R.~H.~Lemmer,
  Annals Phys.\  {\bf 238}, 332 (1995).

\bibitem{Zhu95}
  P.~Zhuang,
  Phys.\ Rev.\  C {\bf 51}, 2256 (1995).

\bibitem{BKRSV95}
  D.~Blaschke, Yu.~L.~Kalinovsky, G.~Roepke, S.~M.~Schmidt and M.~K.~Volkov,
  Phys.\ Rev.\  C {\bf 53}, 2394 (1996)
  [arXiv:nucl-th/9511003].

\bibitem{NBC+96}
  E.~N.~Nikolov, W.~Broniowski, C.~V.~Christov, G.~Ripka and K.~Goeke,
  Nucl.\ Phys.\  A {\bf 608}, 411 (1996)
  [arXiv:hep-ph/9602274].

\bibitem{FB96}
  W.~Florkowski and W.~Broniowski,
  Phys.\ Lett.\  B {\bf 386}, 62 (1996)
  [arXiv:hep-ph/9605315].

\bibitem{OBW00}
  M.~Oertel, M.~Buballa and J.~Wambach,
  Nucl.\ Phys.\  A {\bf 676}, 247 (2000)
  [arXiv:hep-ph/0001239].

\bibitem{Oer00}
  M.~Oertel, M.~Buballa and J.~Wambach,
  Phys.\ Atom.\ Nucl.\  {\bf 64}, 698 (2001)
  [Yad.\ Fiz.\  {\bf 64}, 757 (2001)]
  [arXiv:hep-ph/0008131];
  M.~Oertel,
  PhD thesis, TU Darmstadt (2000)
  [arXiv:hep-ph/0012224].

\bibitem{BBRV08}
  D.~Blaschke, M.~Buballa, A.~E.~Radzhabov and M.~K.~Volkov,
  Yad.\ Fiz.\  {\bf 71}, 2012 (2008)
  [Phys.\ Atom.\ Nucl.\  {\bf 71}, 1981 (2008)]
  [arXiv:0705.0384 [hep-ph]].

\bibitem{Kitazawa:2005mp}
  M.~Kitazawa, T.~Kunihiro and Y.~Nemoto,
  Phys.\ Lett.\  B {\bf 633}, 269 (2006)
  [arXiv:hep-ph/0510167].

\bibitem{PW84}
  R.~D.~Pisarski and F.~Wilczek,
  Phys.\ Rev.\  D {\bf 29}, 338 (1984).

\bibitem{LW60}
  J.~M.~Luttinger and J.~C.~Ward,
  Phys.\ Rev.\  {\bf 118}, 1417 (1960).

\bibitem{BK61}
  G.~Baym and L.~P.~Kadanoff,
  Phys.\ Rev.\  {\bf 124}, 287 (1961).

\bibitem{CJT74}
  J.~M.~Cornwall, R.~Jackiw and E.~Tomboulis,
  Phys.\ Rev.\  D {\bf 10}, 2428 (1974).

\bibitem{Bub05}
  M.~Buballa,
  Phys.\ Rept.\  {\bf 407}, 205 (2005)
  [arXiv:hep-ph/0402234].

\bibitem{Schaefer:2006ds}
  B.~J.~Schaefer and J.~Wambach,
  Phys.\ Rev.\  D {\bf 75}, 085015 (2007)
  [arXiv:hep-ph/0603256].

\bibitem{Schaefer:1999em}
  B.~J.~Schaefer and H.~J.~Pirner,
  Nucl.\ Phys.\  A {\bf 660} (1999) 439
  [arXiv:nucl-th/9903003].

\bibitem{ABC04}
  M.~Alford, J.~Berges and J.~M.~Cheyne,
  Phys.\ Rev.\  D {\bf 70}, 125002 (2004)
  [arXiv:hep-ph/0404059].

\bibitem{GL87}
  J.~Gasser and H.~Leutwyler,
  Phys.\ Lett.\  B {\bf 184}, 83 (1987).

\bibitem{BM61}
  G.~Baym and N.~D. Mermin,
  J.\ Math.\ Phys.\ {bf 2}, 232 (1961).

\bibitem{Bel00}
M.~Le Bellac,
\newblock {\em {Thermal field theory}},
\newblock Cambridge University Press (2000).

\bibitem{AHN01}
  M.~Asakawa, T.~Hatsuda and Y.~Nakahara,
  Prog.\ Part.\ Nucl.\ Phys.\  {\bf 46}, 459 (2001)
  [arXiv:hep-lat/0011040].

\bibitem{WK+02}
  I.~Wetzorke, F.~Karsch, E.~Laermann, P.~Petreczky and S.~Stickan,
  Nucl.\ Phys.\ Proc.\ Suppl.\  {\bf 106}, 510 (2002)
  [arXiv:hep-lat/0110132].

\bibitem{Nic07}
  D.~Nickel,
  Annals Phys.\  {\bf 322}, 1949 (2007)
  [arXiv:hep-ph/0607224].

\bibitem{Bry90}
  R.~K. Bryan,
  Eur.\ Biophys.\ J.\ {\bf 18}, 165 (1990).

\bibitem{Ding:2009ie}
  H.~T.~Ding, O.~Kaczmarek, F.~Karsch, H.~Satz and W.~S\"oldner,
  arXiv:0910.3098 [hep-lat].

\bibitem{Karsch:2009tp}
  F.~Karsch and M.~Kitazawa,
  Phys.\ Rev.\  D {\bf 80}, 056001 (2009)
  [arXiv:0906.3941 [hep-lat]].





\end{thebibliography}

\end{document}